\documentclass[twocolumn,secnumarabic,amssymb, nobibnotes, aps, pra, superscriptaddress]{revtex4-1}
\usepackage{graphicx}
\usepackage{braket}
\usepackage{array}
\usepackage{amsmath}
\usepackage{here}
\usepackage{color}
\usepackage{ulem}
\usepackage{bm}

\begin{document}

\title{Blockade of phonon hopping in trapped ions in the presence of multiple local phonons}%

\author{Ryutaro Ohira}%
\email{u696585a@ecs.osaka-u.ac.jp}
\affiliation{Graduate School of Engineering Science, Osaka University, 1-3 Machikaneyama, Toyonaka, Osaka, Japan}
\author{Shota Kume}%
\affiliation{Graduate School of Engineering Science, Osaka University, 1-3 Machikaneyama, Toyonaka, Osaka, Japan}
\author{Kyoichi Takayama}%
\affiliation{Graduate School of Engineering Science, Osaka University, 1-3 Machikaneyama, Toyonaka, Osaka, Japan}
\author{Silpa Muralidharan}%
\affiliation{Graduate School of Engineering Science, Osaka University, 1-3 Machikaneyama, Toyonaka, Osaka, Japan}
\author{Hiroki Takahashi}%
\affiliation{Center for Quantum Information and Quantum Biology, Institute for Open and Transdisciplinary Research Initiatives, Osaka University, 1-3 Machikaneyama, Toyonaka, Osaka, Japan}
\affiliation{Experimental Quantum Information Physics Unit, Okinawa Institute of Science and Technology Graduate University, 1919-1 Tancha, Onna, Kunigami, Okinawa 904-0495, Japan}
\author{Kenji Toyoda}%
\affiliation{Center for Quantum Information and Quantum Biology, Institute for Open and Transdisciplinary Research Initiatives, Osaka University, 1-3 Machikaneyama, Toyonaka, Osaka, Japan}

\date{\today}

\begin{abstract}
Driving an ion at a motional sideband transition induces the Jaynes--Cummings (JC) interaction. This JC interaction creates an anharmonic ladder of JC eigenstates, resulting in the suppression of phonon hopping due to energy conservation. Here, we realize phonon blockade in the presence of multiple local phonons in a trapped-ion chain. Our work establishes a key technological component for quantum simulation with multiple bosonic particles, which can simulate classically intractable problems.
\end{abstract}


\maketitle

\section*{Introduction}

Trapped ions are a well-isolated and controllable system, providing an ideal platform for quantum information processing (QIP) and quantum simulation. The quantum state of an ion can be expressed by the internal and phonon degrees of freedom. In general, the internal states are used to encode qubits while phonons mediate interactions between the qubits. However, phonons in trapped ions are also a promising resource for QIP and quantum simulation themselves. Phonon states can be optically manipulated by driving motional sideband transitions. Also, because of their high Hilbert-space dimensions, trapped-ion vibrational modes offer large degrees of freedom for use in encoding qubits \cite{1}. 

Phonons in trapped ions can be classified into collective-mode phonons \cite{2,3} and local phonons \cite{4}. When ions are tightly confined, the couplings between the ions are strong, resulting in collective oscillations. However, when the distance between the ions is large, the phonons are localized to each ion in a trapped-ion chain. These local phonons behave as if they were independent particles \cite{5, 6, 7, 8, 9, 10, 11}. Because of their particle-like characteristics and bosonic nature, local phonons can be applied to QIP \cite{12,13} and to quantum simulation of bosonic systems \cite{4, 14} and quantum transport \cite{15,16,17}. In addition, local phonons play an important role in building two-dimensional (2D) trapped-ion systems for QIP and quantum simulation \cite{18, 19, 20}.

A challenge in the experiments with local phonons is unavoidable residual errors in state preparation and detection. In trapped-ion experiments, the desired motional states can be prepared by applying an appropriate pulse sequence. However, to produce a nonclassical state with higher-order Fock states, a relatively long pulse sequence is required. Local phonons constantly hop to other sites even during the state preparation, resulting in an error in state preparation.

A possible approach to reducing the residual errors is phonon blockade \cite{9}. Driving the Jaynes--Cummings (JC) interactions leads to coupling of the internal states and the phonon states. This coupling results in an anharmonic ladder of JC eigenstates, suppressing phonon hopping, i.e., phonon blockade. With implementation of the phonon blockade, state preparation can be realized with fewer hopping-induced errors. This technique can also be applied when the local phonon state is measured with a long pulse sequence.

Realizing state preparation or detection with fewer hopping-induced errors requires implementation of the phonon blockade with multiple local phonons. More specifically, when {\it n} local phonons exist in a trapped-ion chain, the transition $\ket{i-1}\leftrightarrow\ket{i}$ ($i=1,\cdots, n$) in a particular ion site may need to be suppressed. Phonon blockade in the presence of a single local phonon in a three-ion chain has been previously demonstrated \cite{9}. However, to the best of our knowledge, phonon blockade in the presence of multiple phonons has not been reported.

In the present work, we have extended the scheme to multiple phonons and demonstrated two types of blockades at a particular site in the presence of two local phonons in a trapped-ion chain: (1) blockade of the excitation of the first local phonon ($\ket{0}\leftrightarrow\ket{1}$) and (2) blockade of the excitation of the second local phonon ($\ket{1}\leftrightarrow\ket{2}$). Here, we have blocked the excitation of the first local phonon in a targeted ion site in the presence of two local phonons, which is experimentally more challenging than the previous implementation in the sense that the presence of multiple local phonons enhances the flow magnitude of hopping phonons (see the text in \cite{21}). In addition, we have demonstrated blocking the excitation of the second local phonon in a targeted ion site.

\section*{Phonon blockade}

We first describe the principles of phonon blockade. Here, $\it{N}$ ions with mass $\it{m}$ and charge $\it{e}$ form a linear chain in a harmonic potential. The confinement along the $\it{z}$ direction is relatively weak, resulting in large inter-ion distances. Assuming $\hbar=1$ and that the distance between the $\it{i}$th and $\it{j}$th ion is $\it{d_{ij}}$, the Hamiltonian of this system is described in terms of the local phonons as follows \cite{4,22,23}:
\begin{equation}
H_{0} = \sum_{i=\rm 1}^{N}(\omega_{y}+\omega_{i})\hat{a}_{i}^{\dagger}\hat{a}_{i}+\sum_{i<j}^{N}\frac{\it{\kappa_{ij}}}{\rm 2}(\hat{a}_{i}\hat{a}_{j}^{\dagger}+\hat{a}_{i}^{\dagger}\hat{a}_{j}),
\end{equation}
where $\it{\kappa_{ij}}$ and $\omega_{i}$ are the hopping rate between the $\it{i}$th and $\it{j}$th ions and the site-dependent secular frequency shift of the $\it{i}$th ion, respectively:
\begin{equation}
\kappa_{ij}=\frac{e^{\rm 2}}{{\rm 4}\pi\varepsilon_{0}\it{m}\it{d_{ij}}^{\rm 3}\omega_{y}},\,\,\,\,\omega_{i}=-\frac{\rm 1}{\rm 2}\sum_{i\neq j}^{N}{\it{\kappa_{ij}}}.
\end{equation}
Here, $\omega_{y}$ is the secular frequency along the $\it{y}$ direction. In addition, $\hat{a}_{i}$ and $\hat{a}_{i}^{\dagger}$ are the annihilation and creation operators, respectively, of the local phonon mode along the $\it{y}$ direction of the $\it{i}$th ion.

The phonon blockade is based on a quantum nonlinear effect induced by the JC interactions \cite{9}. Illumination of an ion with a laser resonant with a red-sideband induces JC interactions. We assume that 2$\it{g_{i}^{\rm{r}}}$ and $\Delta_{i}^{\rm{r}}$ are the red-sideband Rabi frequency and the detuning from the resonance of the red-sideband transition for the $\it{i}$th ion, respectively. The Hamiltonian of this system after the unitary transformation is performed with $U={\rm exp}(-iH_{1}t)$ ($H_{1}=\sum_{i=\rm 1}^{N}\omega_{y}\hat{a}_{i}^{\dagger}\hat{a}_{i}+\sum_{i=\rm 1}^{N}\Delta_{i}^{\rm{r}}\ket{\uparrow_{i}}\bra{\uparrow_{i}}$) and the rotating wave approximation is described as follows.
\begin{equation}
\begin{split}
  H_{\rm{block}}^{\rm rsb} &= \sum_{i=\rm 1}^{N}\omega_{i}\hat{a}_{i}^{\dagger}\hat{a}_{i}+\sum_{i<j}^{N}\frac{\it{\kappa_{ij}}}{\rm 2}(\hat{a}_{i}\hat{a}_{j}^{\dagger}+\hat{a}_{i}^{\dagger}\hat{a}_{j})\\&+\sum_{i=\rm 1}^{N}\Delta_{i}^{\rm{r}}\ket{\uparrow_{i}}\bra{\uparrow_{i}}+\sum_{i=\rm1}^{N}\it{\it{g_{i}^{\rm r}}}(\hat{a}_{i}\hat{\sigma_{i}}^{+}+\hat{a}_{i}^{\dagger}\hat{\sigma_{i}}^{-}).
\end{split}
\end{equation}
Here, the internal states of an ion are represented as $\ket{\uparrow_{i}}$ and $\ket{\downarrow_{i}}$; $\hat{\sigma_{i}}^{+}=\ket{\uparrow_{i}}\bra{\downarrow_{i}}$ and $\hat{\sigma_{i}}^{-}=\ket{\downarrow_{i}}\bra{\uparrow_{i}}$ are the raising and lowering operators for the $\it{i}$th ion, respectively.

As shown in Fig.\,1(b), the JC interaction results in an anharmonic ladder of JC eigenstates $\ket{\pm,n}$, where $\it{n}$ is the polaritonic excitation number and + and $-$ represent higher and lower branches of eigenstates, respectively. At resonance ($\Delta_{i}^{\rm{r}}=0$), each eigenenergy of the JC Hamiltonian is calculated as
\begin{equation}
E_{i}(n)=n\omega_{i}\pm{g_{i}^{r}}\sqrt{n}.
\end{equation}
The energy difference between two consecutive JC eigenstates is then
\begin{equation}
\begin{split}
\Delta E_{n+1\leftrightarrow n}^{i}&=E_{i}(n+1)-E_{i}(n)\\&=\omega_{i}\pm{g_{i}^{r}}(\sqrt{n+1}-\sqrt{n}).
\end{split}
\end{equation}
This energy difference results in the out-of-resonance coupling between ion-oscillators, preventing local phonons from hopping to the illuminated site.

In our current experimental setup, the relaxation of the blue-sideband Rabi oscillation is empirically known to be longer than that of the red-sideband one. Therefore, in the present experiment, we have implemented phonon blockade using the anti-JC interaction \cite{24}, which can be realized by exciting the ions with lasers resonant with a blue-sideband transition. Therefore, the Hamiltonian is rewritten as 
\begin{equation}
\begin{split}
H_{\rm{block}}^{\rm bsb} &= \sum_{i=\rm 1}^{N}\omega_{i}\hat{a}_{i}^{\dagger}\hat{a}_{i}+\sum_{i<j}^{N}\frac{\it{\kappa_{ij}}}{\rm 2}(\hat{a}_{i}\hat{a}_{j}^{\dagger}+\hat{a}_{i}^{\dagger}\hat{a}_{j})\\&+\sum_{i=\rm 1}^{N}\Delta_{i}^{\rm{b}}\ket{\downarrow_{i}}\bra{\downarrow_{i}}+\sum_{i=\rm1}^{N}\it{\it{g_{i}^{\rm b}}}(\hat{a}_{i}\hat{\sigma_{i}}^{-}+\hat{a}_{i}^{\dagger}\hat{\sigma_{i}}^{+}).
\end{split}
\end{equation}
where 2$\it{g_{i}^{\rm b}}$ and $\Delta_{i}^{\rm{b}}$ are the blue-sideband Rabi frequency and the detuning from the resonance of the blue-sideband transition for the $\it{i}$th ion, respectively. This Hamiltonian is formally equivalent to that in Eq.\,(3), as can be confirmed by interchanging the internal states ($\ket{\downarrow_i}\leftrightarrow\ket{\uparrow_i}$).

\begin{figure}[t]
\centering
  \includegraphics[width=8.0cm]{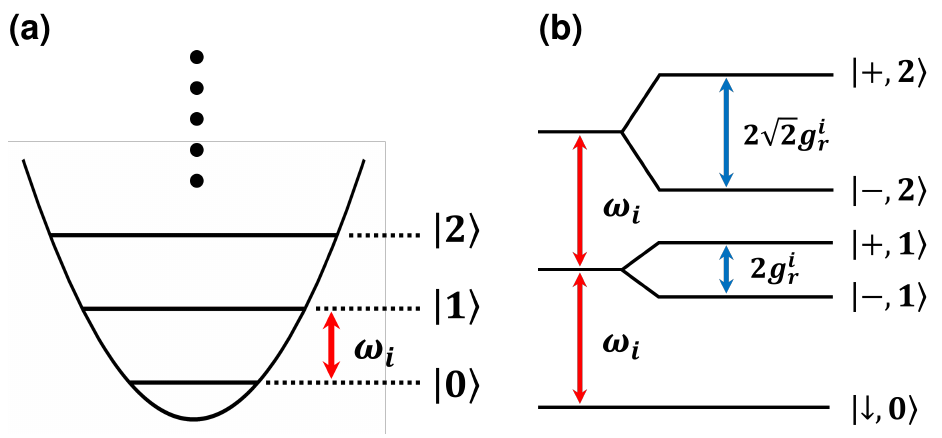}
\caption{\label{fig1} (a) Harmonic ladder of eigenstates of local phonons. (b) Anharmonic ladder of JC eigenstates.}
\end{figure}

\section*{Experimental results}

\subsection*{Blocking the excitation of the first local phonon}

We implement a blockade of the transition between $\ket{0}\leftrightarrow\ket{1}$ in a targeted ion site in the presence of the two local phonons in a trapped-ion chain. In the experiment, two $^{40}{\rm Ca}^{+}$ ions (Ion 1 and Ion 2) are trapped in a linear Paul trap. The secular frequencies along the radial ($\it{x}$ and $\it{y}$) and axial ($\it{z}$) directions are $(\omega_{x}, \omega_{y}, \omega_{z})=$ 2$\pi\times$(3.07, 2.87, 0.11) MHz. The distance between the ions is $\sim$ 24 $\rm{\mu}m$. We use the internal states $\ket{S_{1/2},m_{j}=-1/2}\equiv\ket{\downarrow}$ and $\ket{D_{5/2},m_{j}=-1/2}\equiv\ket{\uparrow}$ to encode the spin states.

The experiment begins with the Doppler cooling of all motional modes and the ground-state cooling of radial motional modes. The narrow quadrupole transition, ${\it S_{\rm 1/2}}$--${\it D_{\rm 5/2}}$, is used for the resolved sideband cooling. The local phonon mode along the {\it y} direction is used. After sideband cooling, two ions are prepared in $\ket{\psi_{\rm init}}=\ket{\psi_{\rm Ion\,1}}\otimes\ket{\psi_{\rm Ion\,2}}=\ket{\uparrow,n_{y}=2}\otimes\ket{\uparrow,n_{y}=0}\equiv\ket{2, 0}$. This state was generated by applying a blue-sideband $\pi$ pulse for the transition between $\ket{\downarrow,n_{y}=0}\leftrightarrow\ket{\uparrow,n_{y}=1}$ to Ion 1 followed by a carrier $\pi$ pulse and a blue-sideband $\pi$ pulse for the transition between $\ket{\downarrow,n_{y}=1}\leftrightarrow\ket{\uparrow,n_{y}=2}$ applied to Ion 1 and Ion 2. A blockade beam was then applied to Ion 2.

\begin{figure}[t]
\centering
  \includegraphics[width=7.5cm]{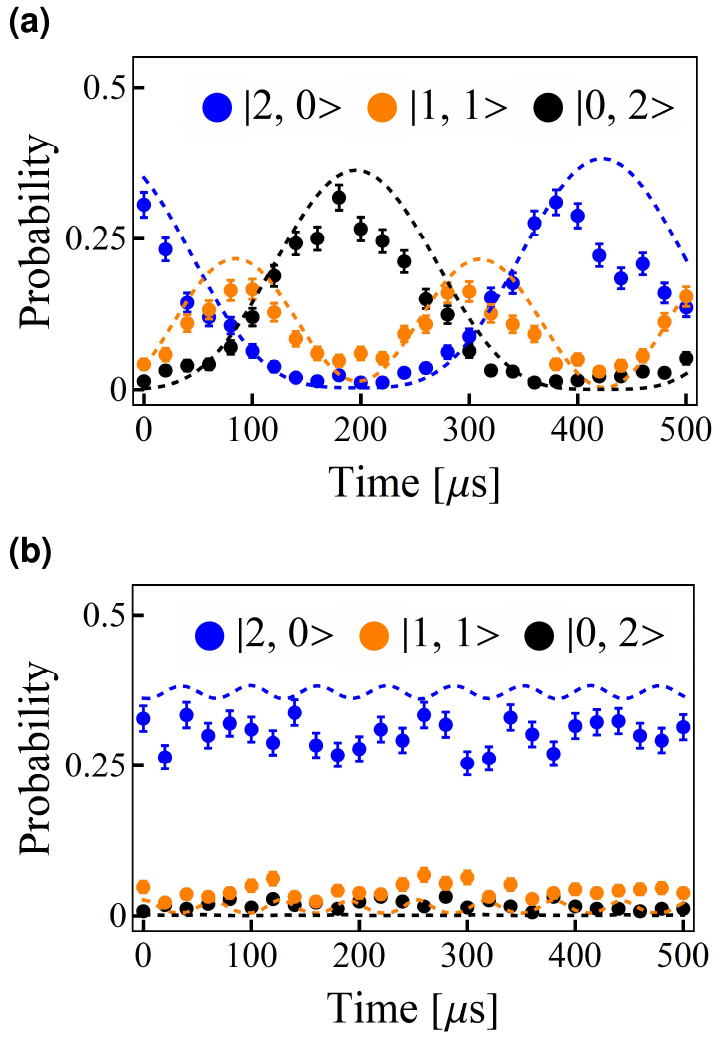}
\caption{\label{fig2} {Experiments on blocking the excitation of the first local phonon. Temporal dynamics of local phonons (a) without blockade and (b) with blockade. The time step between data points is 20 $\rm{\mu}$s. Each data point is an average of 500 measurements. The dashed lines are numerically simulated results. The phonon hopping occurs during the state preparation and the mapping process; therefore, the results start with a derivative.}}
\end{figure}

To observe the multiple phonon dynamics, we use phonon-number-resolving detection \cite{10}. Here, we assume that only the populations in $\ket{\uparrow,n_{y}=0,1,2}$ are finite and neglect the populations in Fock states with higher quantum numbers or in the other internal state. We first apply a composite-pulse sequence at the blue-sideband transition to compensate the Rabi frequency difference between the transitions $\ket{\downarrow,n_{y}=0}\leftrightarrow\ket{\uparrow,n_{y}=1}$ and $\ket{\downarrow,n_{y}=1}\leftrightarrow\ket{\uparrow,n_{y}=2}$. Here, we define the operation for the blue-sideband transition as $R_{\rm BSB}(\theta, \phi)$, where $\theta$ and $\phi$ denote the angle of the qubit rotation for the motional ground state and the rotation axis, respectively. The composite-pulse sequence used in this experiment can then be expressed as 
\begin{equation}
  R_{\rm CP} = R_{\rm BSB}\biggl(\frac{\pi}{2}, 0\biggr)R_{\rm BSB}\biggl(\frac{\pi}{\sqrt{2}}, \frac{\pi}{2}\biggr)R_{\rm BSB}\biggl(\frac{\pi}{2}, 0\biggr).
\end{equation}
Subsequently, a $\pi$ pulse at the carrier transition is applied. The probability amplitude of $\ket{\downarrow,n_{y}=0}$ (originally that of $\ket{\uparrow,n_{y}=0}$) is then transferred to $\ket{D_{5/2},m_{j}=-5/2}\equiv\ket{e_0}$, which has a lifetime of $\sim$ 1 s. Finally, by applying a blue-sideband $\pi$ pulse, the probability amplitude of $\ket{\uparrow,n_{y}=1}$ (originally that of $\ket{\uparrow,n_{y}=2}$) is mapped to the motional ground state $\ket{\downarrow,n_{y}=0}$. At this point, the initial probability amplitudes in $\ket{\uparrow,n_{y}=0,1,2}$ are mapped to the three internal states $\ket{e_{0}}$, $\ket{\uparrow,n_{y}=0}$, $\ket{\downarrow,n_{y}=0}$, respectively. These can be selectively read out by flipping the internal states with carrier pulses and carrying out fluorescence detection. This mapping procedure using a composite-pulse sequence can be performed relatively faster compared with the case of using rapid adiabatic passages for the blue-sideband transition. In this sense, we can suppress the effect of phonon hopping during the detection.

After the mapping process, state-dependent fluorescence detection is performed. If fluorescence is observed, we can conclude that the state was originally in $\ket{\uparrow,n_{y}=2}$; otherwise, the probability amplitude of $\ket{D_{5/2},m_{j}=-5/2}$ is mapped to that of $\ket{\downarrow,n_{y}=0}$ and a  further fluorescence detection is performed. From this result, we can deduce the remaining populations of the original states $\ket{\uparrow,n_{y}=0}$ and $\ket{\uparrow,n_{y}=1}$. Notably, this procedure can be performed independently for Ion 1 and Ion 2. Thus, we can acquire the full information on the populations of the combined motional Fock states (in this case $\ket{2,0}$, $\ket{1,1}$, $\ket{0,2}$).

The observed dynamics of the multiple local phonons without and with a blockade beam are shown in Fig.\,2(a) and Fig.\,2(b), respectively. The measured probabilities for $\ket{2, 0}$, $\ket{1, 1}$, and $\ket{0, 2}$ are shown as functions of the hopping time. The dashed lines in Fig.\,2(a) and Fig.\,2(b) represent the numerically calculated results based on the Hamiltonian in Eq.\,(6), which includes the infidelities of the state preparation and three transitions used in the mapping process. As evident in Fig.\,2, the hopping of the two local phonons is substantially suppressed.

\begin{figure}[t]
\centering
  \includegraphics[width=7.0cm]{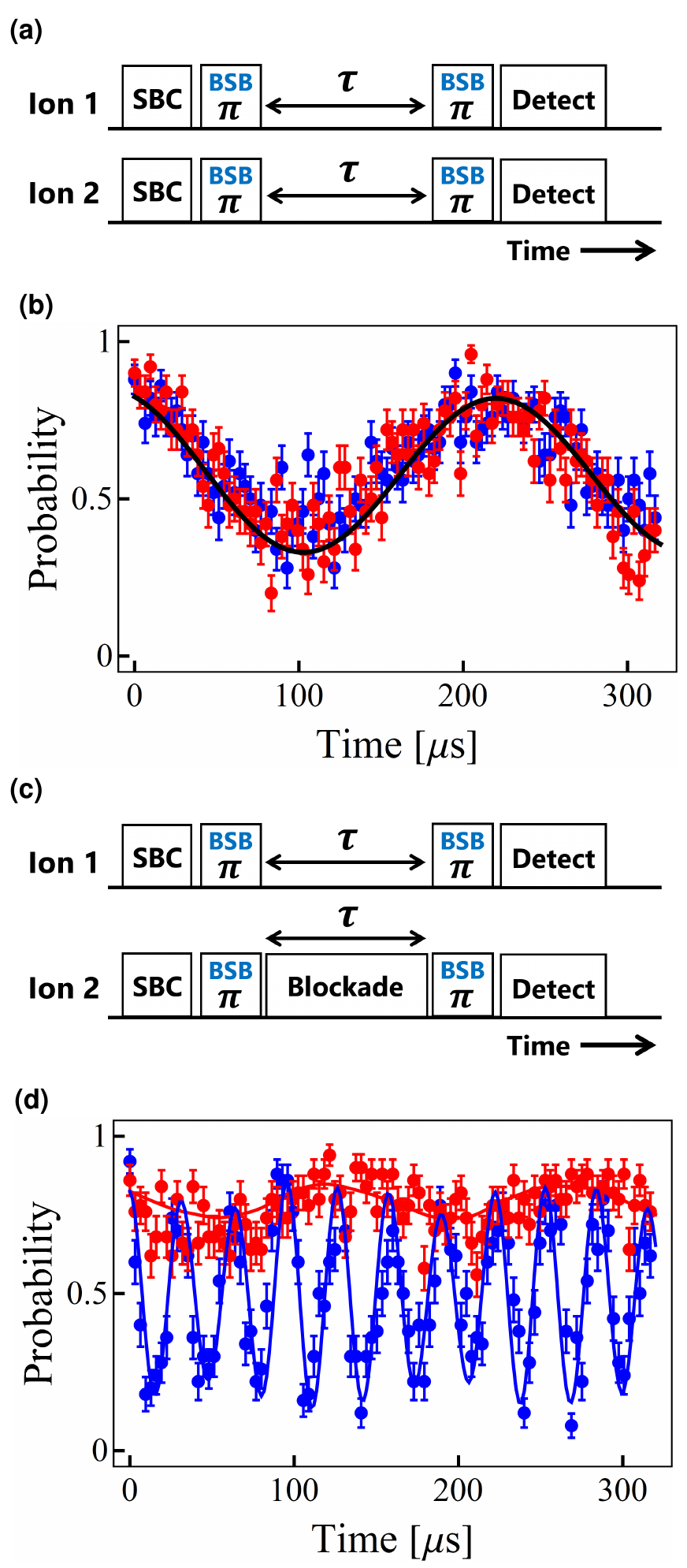}
\caption{\label{fig3} {Experiments on blocking the excitation of the second local phonon using mapping with blue-sideband (BSB) $\pi$ pulses for state analysis. Experimental sequences (a) without blockade and (c) with blockade. We varied hopping (blockade) duration $\tau$ to observe the multiphonon dynamics. Here, SBC represents the sideband cooling. Experimental results (b) without blockade and (d) with blockade. The red and blue circles in the graphs represent the probabilities for Ion 1 and Ion 2, respectively. The solid lines [black line in Fig.\,3(b), red and blue lines in Fig.\,3(d)] represent numerical calculations. Each point is the average of 50 experiments. The error bars represent the statistical uncertainties of 1$\sigma$.}}
\end{figure}

\begin{figure}[t]
\centering
  \includegraphics[width=7.0cm]{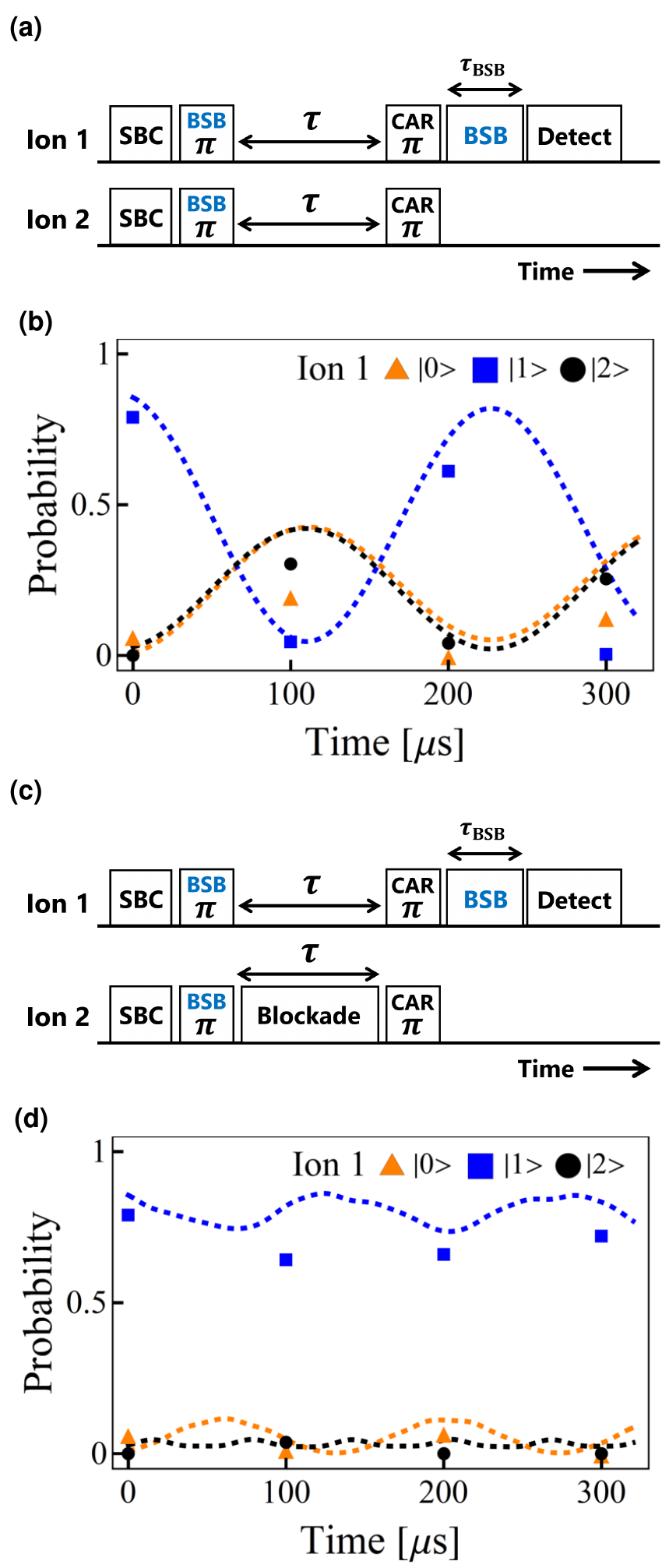}
\caption{\label{fig4} {Experiments on blocking the excitation of the second local phonon using blue-sideband (BSB) Rabi oscillations for state analysis. Experimental sequences (a) without blockade and (c) with blockade. Here, SBC and CAR represent the sideband cooling and the carrier pulse, respectively. To deduce the phonon number distribution, we observe the blue-sideband Rabi oscillation of Ion 1. Experimental results (b) without blockade and (d) with blockade are shown. The probabilities for $\ket{0}$, $\ket{1}$, and $\ket{2}$ obtained from the analysis of the blue-sideband Rabi oscillation are plotted. The dashed lines are numerically calculated results for each Fock state.}}
\end{figure}

\subsection*{Blocking the excitation of the second local phonon}

We next perform an experiment for a blockade of the transition between $\ket{1}\leftrightarrow\ket{2}$ in a targeted ion site. Because of the $\sqrt{n}$ nonlinearity of the JC interaction, as $\it{n}$ increases, $\Delta E_{n+1\leftrightarrow n}^{i}$ decreases. This makes the suppression of the hopping more difficult.

The experimental sequences without and with a blockade beam are shown in Fig.\,3(a) and Fig.\,3(c), respectively. After sideband cooling, both ions are prepared in $\ket{\psi_{\rm init}}=\ket{\psi_{\rm Ion\,1}}\otimes\ket{\psi_{\rm Ion\,2}}=\ket{\uparrow,n_{y}=1}\otimes\ket{\uparrow,n_{y}=1}\equiv\ket{1, 1}$. This state was generated by applying blue-sideband $\pi$ pulses to both ions. Ion 2 was then illuminated with a blockade beam tuned to the blue-sideband transition. After a waiting period equal to the phonon hopping time, each ion was illuminated with a blue-sideband $\pi$ pulse that projected the probability amplitude of $\ket{\uparrow,n_{y}=1}$ onto $\ket{\downarrow,n_{y}=0}$. A 397 nm laser was then used to collect the state-dependent fluorescence.

The experimental results without and with a blockade beam are shown in Fig.\,3(b) and Fig.\,3(d), respectively. In these experiments, the quantum states of the ions before the detection can be expressed as follows:
\begin{equation}
   \left\{
   \begin{aligned}
       \ket{\psi_{\rm Ion\,1}(t)} = \sum_{n=0}^{2}c_{\downarrow,n}^{\rm Ion\,1}\ket{\downarrow,n}+\sum_{n=0}^{1}c_{\uparrow,n}^{\rm Ion\,1}\ket{\uparrow,n} \\
       \ket{\psi_{\rm Ion\,2}(t)} = \sum_{n=0}^{2}c_{\downarrow,n}^{\rm Ion\,2}\ket{\downarrow,n}+\sum_{n=0}^{1}c_{\uparrow,n}^{\rm Ion\,2}\ket{\uparrow,n},
   \end{aligned}
   \right.
\end{equation}
where $c_{\downarrow,n}^{\rm Ion\,1(Ion\,2)}$ and $c_{\downarrow,n}^{\rm Ion\,1(Ion\,2)}$ are probability amplitudes for each state. The probability that we obtain in this experiment is
\begin{equation}
  P_{\rm{Ion\,1(Ion\,2)}} = |\braket{\downarrow|\psi_{\rm Ion\,1(Ion\,2)}(t)}|^2.
\end{equation}

The experiment without a blockade beam [Fig.\,3(b)] corresponds to the Hong--Ou--Mandel experiment \cite{8,10,25}. The solid black line in Fig.\,3(b) represents numerically calculated results for Ion 1 and Ion 2. In the experiment with a blockade beam [Fig.\,3(d)], the state of Ion 1 remains largely constant, whereas Ion 2 experiences Rabi flopping caused by the blockade beam. The solid red and blue lines in Fig.\,3(d) represent numerical calculation results for Ion 1 and Ion 2, respectively. Each data point in both graphs is the average of 50 measurements. In the numerical calculation, we include the imperfection of the sideband cooling and the infidelities of the carrier and blue-sideband Rabi oscillation.

The results in Fig.\,3(d) show a strong indication that the transition from $\ket{1}$ to $\ket{0}$ or $\ket{2}$ in Ion 1 is suppressed. However, to confirm that the blockade is working, we also need to deduce the phonon number distribution of each ion. To extract the phonon number distribution, we perform another experiment shown in Fig.\,4(a) (without blockade) and Fig.\,4(c) (with blockade). Here, the analysis of the blue-sideband Rabi oscillation is conducted. After waiting for the hopping time, we apply a carrier $\pi$ pulse to each ion followed by measuring the blue-sideband Rabi oscillation of Ion 1.

The results for the experiments without and with a blockade beam are given in Fig.\,4(b) and Fig.\,4(d), respectively. By fitting the blue-sideband Rabi oscillation, we extract the phonon number distribution \cite{24}. The dashed lines in each graph are numerically calculated results for each motional Fock state. On the basis of the results in Fig.\,4(b) and (d), we conclude that the transitions from $\ket{1,1}$ to $\ket{0,2}$ and $\ket{2,0}$ are blocked.

\section*{Discussion}

The contrast of the results is limited by the infidelities of the state preparation and the mapping process, in particular that of the blue-sideband $\pi$ pulses ($0.1\sim0.15$). In the case of the state preparation, the average motional number along the radial direction is approximately 0.04. To confirm the validity of the experimental data, we have performed a numerical calculation for each experiment. For the numerical simulation, we have included infidelities of the state preparation and $\pi$ pulses for each transition into a Liouville equation as Lindblad-type relaxation.

In the present phonon blockade method, we apply a blockade beam resonant with the blue-sideband transition, thereby driving the transition between $\ket{\downarrow,0}\leftrightarrow\ket{\uparrow,1}$ of the targeted ion. This approach can cause difficulties in local phonon operations. Here, we propose a method to avoid this Rabi flopping. The evolution of the Bloch vector $\bm{R}$ on the Bloch sphere can be represented as
\begin{equation}
  \frac{d\bm{R}}{dt} = \bm{K}\times\bm{R}.
\end{equation}
Eq.\,(9) shows that the Bloch vector precesses around the vector $\bm{K}$. Therefore, the sideband Rabi flopping can be avoided by satisfying $\bm{K}\times\bm{R}=0$; i.e., the Bloch vector is either parallel or antiparallel to the vector $\bm{K}$. We can realize this situation by using adiabatic passage over the sideband transition \cite{26,27,28}. Assuming that the quantum state of the targeted ion is $\ket{\psi}=\ket{\uparrow,n}$, we then prepare the ion in the superposition of $\ket{\downarrow,n-1}$ and $\ket{\uparrow,n}$ using adiabatic passage over the blue-sideband transition. The condition $\bm{K}\times\bm{R}=0$ can be satisfied by stopping the sweep of the power and detuning of the laser at the halfway point of the adiabatic process. However, the speed of the adiabatic passage is usually comparable to or slower than the hopping period. To mitigate this problem, we propose using transitionless quantum driving to accelerate the adiabatic process \cite{29,30,31,32}. By implementing the transitionless quantum driving into the aforementioned idea, we realize the steady Bloch vector during the blockade. In addition, the adiabatic passage can eliminate the phonon-number-dependence on the sideband Rabi oscillation; thus, confining an arbitrary superposition of Fock states or a nonclassical state, such as a coherent state and squeezed state, to a limited region in a trapped ion chain might be possible.

\section*{Conclusions}
We have demonstrated phonon blockade in the presence of two local phonons in a trapped-ion chain. Our work is an important step toward the realization of a large-scale quantum simulator with multiple phonons.

\section*{Acknowledgments}
This work was supported by MEXT Quantum Leap Flagship Program (MEXT Q-LEAP) Grant Number JPMXS0118067477.


\begin{thebibliography}{99}
\bibitem{21}For the hopping Hamiltonian for two ions ($\psi=\psi_{\rm Ion\,1}\otimes\psi_{\rm Ion\,2}\equiv\ket{n_{\rm Ion\,1},n_{\rm Ion\,2}}$) $H_{\rm Hop}=\frac{\kappa_{12}}{2}\,(a_1a_2^\dagger+{a_1^\dagger}a_2)$, the rate for each transition from $\ket{n,0}$ to $\ket{n-1,1}$ and from $\ket{1,0}$ to $\ket{0,1}$ can be calculated as $\bra{n-1,1}H_{\rm Hop}\ket{n,0}=\frac{\kappa_{12}}{2}\sqrt{n}$ and $\bra{0,1}H_{\rm Hop}\ket{1,0}=\frac{\kappa_{12}}{2}$, respectively. Therfore, the rate for $\ket{n,0}\rightarrow\ket{n-1,1}$ becomes faster than that for $\ket{1,0}\rightarrow\ket{0,1}$ by the factor $\sqrt{n}$.
\end{thebibliography}
\end{document}